\documentclass[a4paper]{jpconf}
\usepackage{graphicx}
\usepackage{amsmath}
\begin{document}
\title{Observing the Spontaneous Breakdown of Unitarity}

\author{Jasper van Wezel}

\address{Theory of Condensed Matter, 19 JJ Thomson Avenue, Cambridge CB3 0HE, UK}

\ead{physics@jvanwezel.com}

\begin{abstract}
During the past decade, the experimental development of being able to create ever larger and heavier quantum superpositions has brought the discussion of the connection between microscopic quantum mechanics and macroscopic classical physics back to the forefront of physical research. Under equilibrium conditions this connection is in fact well understood in terms of the mechanism of spontaneous symmetry breaking, while the emergence of classical dynamics can be described within an ensemble averaged description in terms of decoherence. The remaining realm of individual-state quantum dynamics in the thermodynamic limit was addressed in a recent paper proposing that the unitarity of quantum mechanical time evolution in macroscopic objects may be susceptible to a spontaneous breakdown. Here we will discuss the implications of this theory of spontaneous unitarity breaking for the modern experiments involving truely macroscopic Schr\"odinger cat states.
\end{abstract}

\section{Introduction}
The connection between the quantum behaviour of microscopic particles and the
classical laws that govern macroscopic bodies has puzzled generations of
physicists~\cite{Jammer:old}. In the present day, the quest for the quantum computer and our
ability to directly manipulate the quantum properties of ever larger objects
have brought the micro-macro connection back to the front lines of fundamental
research. It has now become possible to routinely create Schr\"odinger cat states using for example a supercurrent consisting of up to $10^{11}$ electrons~\cite{Lukens:flux1,Mooij:flux03}, or a molecule whose De Broglie wavelength is much shorter than its own diameter~\cite{Zeilinger99}. In a recent paper, Marshall et al. propose to extend the reach of these types of experiments over several orders of magnitude, by superposing a micron-sized mirror over two distinct spatial positions~\cite{Marshall03}. The possibility of realizing such macroscopic superpositions in the laboratory offers a unique opportunity to experimentally access the question of how the quantum mechanics of microscopic scales is connected to the classical description of macroscopic objects.

It has recently been proposed that classical dynamics may result from a spontaneous breakdown of the unitarity of quantum mechanical time evolution in macroscopic objects~\cite{vanWezel:DSSB}. The proposed mechanism is a straightforward extension of the equilibrium theory of spontaneous symmetry breaking to the case of quantum dynamics. It is applicable to the dynamics of a single quantum state, and thus augments the predictions of decoherence, which necessarily involve an average over all possible states of the unobservable environment~\cite{vanWezel:DSSB,Bassi00,Adler}. 
%
%
%

Here, we will discuss the possible implications of the presence of spontaneous unitarity breaking for modern, truely macroscopic, Schr\"odinger cat experiments such as the one proposed by Marshall et al.
We will first give a brief overview of the descriptions of spontaneous symmetry breaking and spontaneous unitarity breaking and comment on how they are related. We will then discuss the requirements for realizing a spontaneous breakdown of unitarity, and discuss how it could influence a typical Schr\"odinger cat experiment.

\section{Spontaneous Symmetry Breaking}
%
The heart of the workings of spontaneous symmetry breaking lies in the fact that every many-particle Hamiltonian which possesses a continuous symmetry that is unbroken in its ground state, gives rise to a tower of low-energy states called the thin spectrum~\cite{vanWezel07,vanWezel05}. The states in this thin spectrum represent global (infinite wavelength) excitations that can be seen as the centre of mass properties of the collective system~\cite{vanWezel07,Birol07}. As an example, one could think of the rotationally invariant Hamiltonian for a quantum (antiferro)magnet, in which the thin spectrum consists of the eigenstates of the total spin operator acting on the combined spin of all magnetic moments in the entire magnet. The non-degenerate ground state of such a model can be easily shown to be the isotropic singlet state with zero total spin~\cite{Lieb62}. In order to explicitly break the symmetry, one has to introduce a coupling between different thin spectrum states so that the ground state in the presence of that symmetry breaking field has a well-defined order parameter:
\begin{align}
\hat{H}_{SB} = \hat{H}_{0} + \hat{O}.
\label{Hssb}
\end{align}
Here $\hat{H}_0$ is the symmetric Hamiltonian and $\hat{O}$ is the symmetry breaking, or order parameter field which couples different thin spectrum states. In the case of the antiferromagnet, the symmetry breaking field would consist of a staggered magnetic field which points up on one of the sublattices and down on the other. The resulting classical antiferromagnetic ground state is described by a wavepacket built up out of many different total spin states.

The crucial observation is now that the strength of the field needed to give rise to a fully ordered ground state depends on the total number of particles, $N$, in the system~\cite{Anderson:SolidState}. Because the energy separation between two consecutive thin spectrum states typically scales as $1/N$, the field strength necessary to explicitly break the symmetry decreases with system size. In the thermodynamic limit (where $N \to \infty$) all of the thin spectrum states collapse onto the ground state to form a degenerate continuum of states. Within this continuum even an infinitesimally small symmetry breaking field is enough stabilise a fully ordered, symmetry broken state. 

\section{Spontaneous Unitarity Breaking}
The extension of the equilibrium theory of spontaneous symmetry breaking to the dynamic case of spontaneous unitarity breaking~\cite{vanWezel:DSSB} follows naturally from the realisation that within quantum mechanics, the Hamiltonian operator plays a dual role. On the one hand its eigenstates form the set of possible stable states that an isolated quantum system can occupy, while on the other hand it also serves as the generator for the time evolution of a general initial state. The theory of spontaneous symmetry breaking discussed in the previous section addresses the fate of the Hamiltonian's first role in the thermodynamic limit. In that limit it becomes possible for symmetry broken states that are not eigenstates of $\hat{H}_0$ to nonetheless be realized. A similar fate also affects the Hamiltonian's second role: the unitary time translation symmetry that is generated by the Hermitian Hamiltonian operator $\hat{H}_0$ can spontaneously break down in the thermodynamic limit. Closely following the standard description of equilibrium symmetry breaking, we can explicitly break time translation symmetry by introducing a small unitarity breaking field into the generator of time evolution~\cite{vanWezel:DSSB}:
\begin{align}
\hat{U}(t) = \exp \left(- \frac{i}{\hbar} t \left[\hat{H}_0 - i \hat{O} \right] \right).
\label{Hdssb}
\end{align}
Here $\hat{H}_0$ is again the symmetric Hamiltonian and $\hat{U}(t)$ is the time evolution operator. The field $\hat{O}$ is the same order parameter field that was introduced in equation~\eqref{Hssb}. Its introduction here in the form $i\hat{O}$ makes the operator for time evolution explicitly non-unitary. The quantum time evolution that is generated by a non-unitary $\hat{U}(t)$ does not in general conserve energy. However, because the operator $\hat{O}$ only couples states within the thin spectrum (whose energies all scale as $1/N$), conservation of energy is in this case automatically restored in the thermodynamic limit.

The unitarity broken time evolution implied by the operator $\hat{U}(t)$ can give rise to two possible types of behaviour, depending on which initial state it is applied to~\cite{vanWezel:DSSB}. If the initial state has a finite overlap with the symmetry broken state favoured by the operator $\hat{O}$, then this symmetry broken state will be strongly amplified during the time evolution, and it will completely dominate the final wavefunction within a timescale $\tau \propto \hbar/No$, where $o$ represents the strength of the field $\hat{O}$. If on the other hand the initial state has zero overlap with the favoured symmetry broken state (for example because it is in a differently ordered state), then the system must first employ the unitary part of the time evolution operator to explore its phase space. Only after a timescale proportional to the ergodic time (which goes to infinity in the thermodynamic limit), does the system again have a finite overlap with the amplified state. The result is thus that in the thermodynamic limit even an infinitesimally small non-unitary field suffices to almost instantaneously reduce a non-classical (i.e. not explicitly symmetry broken) state to a fully ordered classical state~\cite{vanWezel:DSSB}, while it does not have any discernable influence on states that already possess a broken symmetry.

\section{The Unitarity Breaking Field}
The unitarity breaking field that we introduced in equation~\eqref{Hdssb}, is clearly not a normal quantum field. If it exists, it must come from a part of physics outside of quantum mechanics, which can be effectively described on some scale by the introduction of a small non-unitary correction to the usual quantum theory. The precise form of this correction in fact is of very little consequence for the workings of spontaneous unitarity breaking. Already in the equilibrium case of equation~\eqref{Hssb}, the 'usual' symmetry breaking field~$\hat{O}$ could be formed by only an infinitesimal component of some physical perturbation to the ideal system~$\hat{H}_0$. That is, even if a few magnetic impurities in a real antiferromagnet do not form an exact staggered magnetic field over all sites of the lattice, they can still stabilise the antiferromagnetic symmetry broken state, because any component of their combined influence that does look like a staggered field is amplified by a factor $N$ through the influence of the thin spectrum states. For spontaneous unitarity breaking the precise form and strength of the non-unitary perturbation are likewise unimportant, because any component that agrees with the field~$i\hat{O}$ is amplified through the thin spectrum, so that its rate of influence is sped up by a factor $N$.

The idea that quantum mechanics may need to be corrected by a non-unitary influence on some scale, is not a surprise at all. In fact, the explicit need for the strict linearity of quantum mechanics to break down before macroscopic scales are reached can be shown on quite general grounds~\cite{Bassi00}. Although quantum mechanics is a strictly unitary theory, other areas of physics do not have this constraint. General Relativity for example is a notably non-unitary theory, whose possible influence on the quantum mechanics of mesoscopic systems has been considered before~\cite{Diosi,Penrose:96,VanWezel:penrose}. 

A main motivation for doing macroscopic Schr\"odinger cat experiments, such as the one proposed by Marshall et al.~\cite{Marshall03}, is to find out at which scale the non-unitary corrections to quantum mechanics become important and to isolate their source. The theory of spontaneous unitarity breaking does not address the possible sources of non-unitary fields, but it does predict what the influence of a non-unitary correction, regardless of its source, must be. If the object is large enough for the timescale $\tau \propto \hbar/No$ to be unobservably short, the only possible outcome of the time evolution generated via equation~\eqref{Hdssb} is the seemingly instantaneous reduction of a quantum superposition to just a single classical state. The probability for which classical state is selected turns out to be given by Born's rule~\cite{vanWezel:DSSB}. The influence of the non-unitary field on microscopic objects on the other hand, is neglicably small, and their time evolution must correspondingly always look purely quantum mechanical. The interesting regime of mesoscopic objects, which are just large enough to be noticebly influenced by corrections to quantum mechanics, while still being small enough to keep the non-unitary time evolution within the experimentally accessible timescales, is precisely the regime that has not been explored yet, and that the new generation of Schr\"odinger cat experiments may just begin to investigate~\cite{Penrose:96}.

\section{Conclusions}
We have shown that the same properties which allow a macroscopic object under equilibrium conditions to be in a state that does not obey the symmetry of its governing Hamiltonian, also allow such a macroscopic object to escape the unitarity of the quantum mechanical time evolution generated by that Hamiltonian. Through this process of spontaneous unitarity breaking, objects in the thermodynamic limit become almost infinitely sensitive to any infinitessimal correction to the theory of quantum mechanics. The central result of the present work is that because the thin spectrum in macroscopic objects only amplifies those non-unitarities that couple directly to the order parameter, the resulting non unitary time evolution of a general initial state does not depend on details of the non-unitary correction. If the initial state is a Schr\"odinger cat like superposition of macroscopically distinct states, the unitarity-broken evolution will eventually automatically give rise to Born's rule.

To experimentally access the non-unitary time evolution and isolate its source, it will be necessary to create a Schr\"odinger cat state using mesoscopic objects. For such superposition states the rule of thumb deduced from spontaneous unitarity breaking can be formulated as: 'If an object is large enough to display a spontaneously broken symmetry, then its order parameter cannot be forced into a stable superposition.' Thus an experiment such as the one proposed by Marshall et al. in which a clearly localized, mesoscopic mirror is brought into a spatial superposition, should have a good chance of observing spontaneously broken unitarity.

%
%

\section*{References}
%
\providecommand{\newblock}{}

\end{document}